\pdfoutput=1

\documentclass{article}

\usepackage{microtype}
\usepackage{graphicx}
\usepackage{subfigure}
\usepackage{booktabs} 

\usepackage{hyperref}
\usepackage{url}
\usepackage{graphicx}
\usepackage{makecell}
\usepackage{multirow}
\usepackage{enumitem}
\usepackage{bbm}
\usepackage{color,xcolor}
\usepackage{booktabs}
\usepackage{lineno}
\usepackage{array}
\usepackage{ulem}
\usepackage{makecell}
\usepackage{multicol}
\usepackage{amsmath,amsthm,amsfonts,amssymb,bm}
\usepackage{stfloats}
\usepackage[font=small,labelfont=bf]{caption}
\usepackage{tabulary}
\newcolumntype{C}[1]{>{\centering\arraybackslash}p{#1}}
\newcolumntype{L}[1]{>{\arraybackslash}p{#1}}

\usepackage[accepted]{icml2021}

\begin{document}

\twocolumn[
\icmltitle{AlphaDesign: A graph protein design method and benchmark on AlphaFoldDB}
\vspace{-4mm}
\icmlsetsymbol{equal}{*}


\begin{icmlauthorlist}
\icmlauthor{Zhangyang Gao}{equal,to,goo}
\icmlauthor{Cheng Tan}{equal,to,goo}
\icmlauthor{Stan Z. Li}{to}
\end{icmlauthorlist}

\icmlaffiliation{to}{AI Research and Innovation Lab, Westlake University}
\icmlaffiliation{goo}{Zhejiang University}

\icmlcorrespondingauthor{Stan Z. Li}{Stan.ZQ.Li@westlake.edu.cn}

\icmlkeywords{Deep Learning, Protein design}

\vskip 0.1in
]



\printAffiliationsAndNotice{\icmlEqualContribution} 

\begin{abstract}
    While DeepMind has tentatively solved protein folding, its inverse problem -- protein design which predicts protein sequences from their 3D structures -- still faces significant challenges. Particularly, the lack of large-scale standardized benchmark and poor accuray hinder the research progress. In order to standardize comparisons and draw more research interest, we use AlphaFold DB, one of the world's largest protein structure databases, to establish a new graph-based benchmark -- AlphaDesign. Based on AlphaDesign, we propose a new method called ADesign to improve accuracy by introducing protein angles as new features, using a simplified graph transformer encoder (SGT), and proposing a confidence-aware protein decoder (CPD). Meanwhile, SGT and CPD also improve model efficiency by simplifying the training and testing procedures. Experiments show that ADesign significantly outperforms previous graph models, e.g., the average accuracy is improved by 8\%, and the inference speed is 40+ times faster than before.

\end{abstract}

\vspace{-7mm}
\section{Introduction}

As "life machines," proteins play vital roles in almost all cellular processes, such as transcription, translation, signaling, and cell cycle control. Understanding the relationship between protein structures and their sequences brings significant scientific impacts and social benefits in many fields, such as bioenergy, medicine, and agriculture \citep{huo2011conversion, williams2019plasma}. While AlphaFold2 \citep{jumper2021highly} has tentatively solved protein folding from 1D sequences to 3D structures, its reverse problem, i.e., protein design raised by \citep{pabo1983molecular} that aims to predict amino acid sequences from known 3D structures, has fewer breakthroughs in the ML community. The main reasons hindering the research progress include: (1) The lack of large-scale standardized benchmarks; (2) The difficulty in improving protein design accuracy; (3) Many methods are neither efficient nor open source. Therefore, a crucial question grips our heart: How to benchmark protein design and develop an effective and efficient open-sourced method?

Previous benchmarks may suffer from biased testing and unfair comparisons. Since SPIN \citep{li2014direct} introduced the TS50 consisting of 50 native structures, it has beening served as a common test set for evaluating different methods \citep{o2018spin2,wang2018computational,chen2019improve,jing2020learning,zhang2020prodconn,qi2020densecpd,strokach2020fast}. However, such a few proteins do not cover the vast protein space and are more likely to lead to biased tests. Besides, there are no canonical training and validation sets, which means that different methods may use various training sets. If the training data is inconsistent, how can we convince researchers that the performance gain comes from the difference of methods rather than data? Especially when the test set is small, adding training samples that match the test set distribution may dramatic performance fluctuations. For fair comparisions, how to establish a large-scale standardized benchmark?

Extracting expressive residue representations is a key challenge for accurate protein design, where both sequential and structural properties need to be conisdered. For general 3D points, structural features should be rotationally and translationally invariant in the classification task. Regarding proteins, we should further consider the regular structure, number, and order of amino acids. Previous studies \citep{o2018spin2,wang2018computational,ingraham2019generative,jing2020learning} may have overlooked some important protein features and better model designs; thus, none of them exceed 50\% accuracy except DenseCPD \citep{qi2020densecpd}. For higher accuracy, how to construct better protein features and neural models to learn expressive residue representations?

Improving the model efficiency is necessary for rapid iteration of researches and applications. Current advanced GNN and CNN methods have severe speed defects due to the sequential prediction paradigm. For example, GraphTrans \citep{ingraham2019generative} and GVP \citep{jing2020learning} predict residues one by one during inference rather than in parallel, which means that it calls the model multiple times to get the entire protein sequence. Moreover, DenseCPD \citep{qi2020densecpd} takes 7 minutes to predict a 120-length protein on their server \footnote{\url{http://protein.org.cn/densecpd.html}}. How can we improve the model efficiency while ensuring accuracy?


To address these problems, we establish a new protein design benchmark -- AlphaDesign, and develop a graph model called ADesign to achieve SOTA accuracy and efficiency. Firstly, 
AlphaDesign compares various graph models on consistent training, validation, and testing sets, where all these datasets come from the AlphaFold Protein Structure Database \citep{varadi2021alphafold}. In contrast to previous studies  \citep{ingraham2019generative,jing2020learning} that use limited-length proteins and do not distinguish species, we extend the experimental setups to length-free and species-aware. Secondly, we improve the model accuracy by introducing protein angles as new features, using a simplified graph transformer encoder (SGT), and proposing a constraint-aware protein decoder (CPD). Thirdly, SGT and CPD also improve model efficiency by simplifying the training and testing procedures. Experiments show that ADesign significantly outperforms previous methods in both accuracy (+8\%) and efficiency (40+ times faster than before).

\section{Related work}
We focus on structure-based protein design, and the mainstream approaches can be categorized into three groups, i.e., MLP-based, CNN-based, and GNN-based methods. Some terms need to be explained in advance: we call amino acids in the protein as residues, and accuracy is an indicator of how well the residues predict.



\paragraph{Problem definition} The structure-based protein design aims to find the amino acids sequence $\mathcal{S} = \{s_i: 1 \leq i \leq n\}$ that folds into a known 3D structure $\mathcal{X}=\{\boldsymbol{x}_{i} \in \mathbb{R}^3: 1 \leq i \leq n \}$, where $n$ is the number of residues and the natural proteins are composed by 20 types of amino acids, i.e., $1\leq s_i \leq 20$. Formally, that is to learn a function $\mathcal{F}_{\theta}$:
\begin{align}
    \label{eq:protein_dedign}
    \mathcal{F}_{\theta}: \mathcal{X} \mapsto \mathcal{S}.
\end{align}
Because homologous proteins always share similar structures \citep{pearson2005limits}, the problem itself is underdetermined, i.e., the valid amino acid sequence may not be unique. In addition, the need to consider both 1D sequential and 3D structural information further increases the difficulty of algorithm design. As a result, the published algorithms are still not accurate enough.



\vspace{-2mm}
\paragraph{MLP-based models} These methods use \textbf{m}ulti-\textbf{l}ayer \textbf{p}erceptron (MLP) to predict the type of each residue. The MLP outputs the probability of 20 amino acids for each residue, and the input feature construction is the main difference between various methods. SPIN \citep{li2014direct} integrates torsion angles ($\phi$ and $\psi$), fragment-derived sequence profiles and structure-derived energy profiles to predict protein sequences. SPIN2 \citep{o2018spin2} adds backbone angles ($\theta$ and $\tau$), local concat number and neighborhood distance to improve the accuracy from 30\% to 34\%. \citep{wang2018computational} uses backbone dihedrals ($\phi$, $\psi$ and $\omega$), sovlent accessible surface area of backbone atoms ($C_{\alpha}, N, C,$ and $O$), secondary structure types (helix, sheet, loop), $C_{\alpha}-C_{\alpha}$ distance and unit direction vectors of $C_{\alpha}-C_{\alpha}$, $C_{\alpha}-N$ and $C_{\alpha}-C$, which acheives 33.0\% accuracy on 50 test proteins. The MLP methods have a high inference speed, but their accuracy is relatively low due to the partial considering of structural information. The complex feature engineering requires multiple databases and computational tools, limiting the widespread usage.

\vspace{-2mm}
\paragraph{CNN-based models} CNN methods extract protein features directly from the 3D structure, which can be further classified as 2D CNN-based and 3D CNN-based, with the latter achieving better results. The 2D CNN-based SPROF \citep{chen2019improve} extracts structural features from the distance matrix, and improves the accuracy to 39.8\%. In contrast, 3D CNN-based methods extract residue features from the atom distribution in a three-dimensional grid box. For each residue, the atomic density distribution is computed after being translated and rotated to a standard position so that the model can learn translation and rotation invariant features. ProDCoNN \citep{zhang2020prodconn} designs a nine-layer 3D CNN to predict the corresponding residues at each position, which uses multi-scale convolution kernels and achieves 42.2\% accuracy. DenseCPD \citep{qi2020densecpd} further uses the DensetNet architecture \citep{huang2017densely} to boost the accuracy to 55.53\%. Although 3DCNN-based models improve accuracy, their inference is slow, probably because they require separate pre-processing and prediction for each residue. In addition, none of the above 3D CNN models are open source.


\vspace{-2mm}
\paragraph{Graph-based models} Graph methods represent the 3D structure as a $k$-NN graph, then use graph neural networks \citep{defferrard2016convolutional,kipf2016semi,velivckovic2017graph,zhou2020graph,zhang2020deep} to extract residue features while considering structural constraints. The protein graph encodes the residue information in node vectors and constructs edges and edge features between neighboring residues. GraphTrans \citep{ingraham2019generative} achieves 36.4\% accuracy using graph attention and auto-regression generation strategies. GVP \citep{jing2020learning} increases the accuracy to 40.2\% by proposing the geometric vector perceptron, which learns both scalar and vector features in an equivariant and invariant manner with respect to rotations and reflections. Another related work is ProteinSolver \citep{strokach2020fast}, but it was mainly developed for scenarios where partial sequences are known and does not report results on standard datasets. Firstly, since both GraphTrans and GVP use rotated, translation-invariant features, there is no need to rotate each residue separately as in CNN, thus improving the training efficiency. However, during testing, GraphTrans and GVP use an autoregressive mechanism to generate residuals one by one rather than in parallel, which weakens their advantage in inference speed. Secondly, the well-exploited structural information helps GNN obtain higher accuracy than MLPs. In summary, GNN can achieve a good balance between efficiency and accuracy.

\begin{figure*}[t]
    \centering
    \includegraphics[width=6.5in]{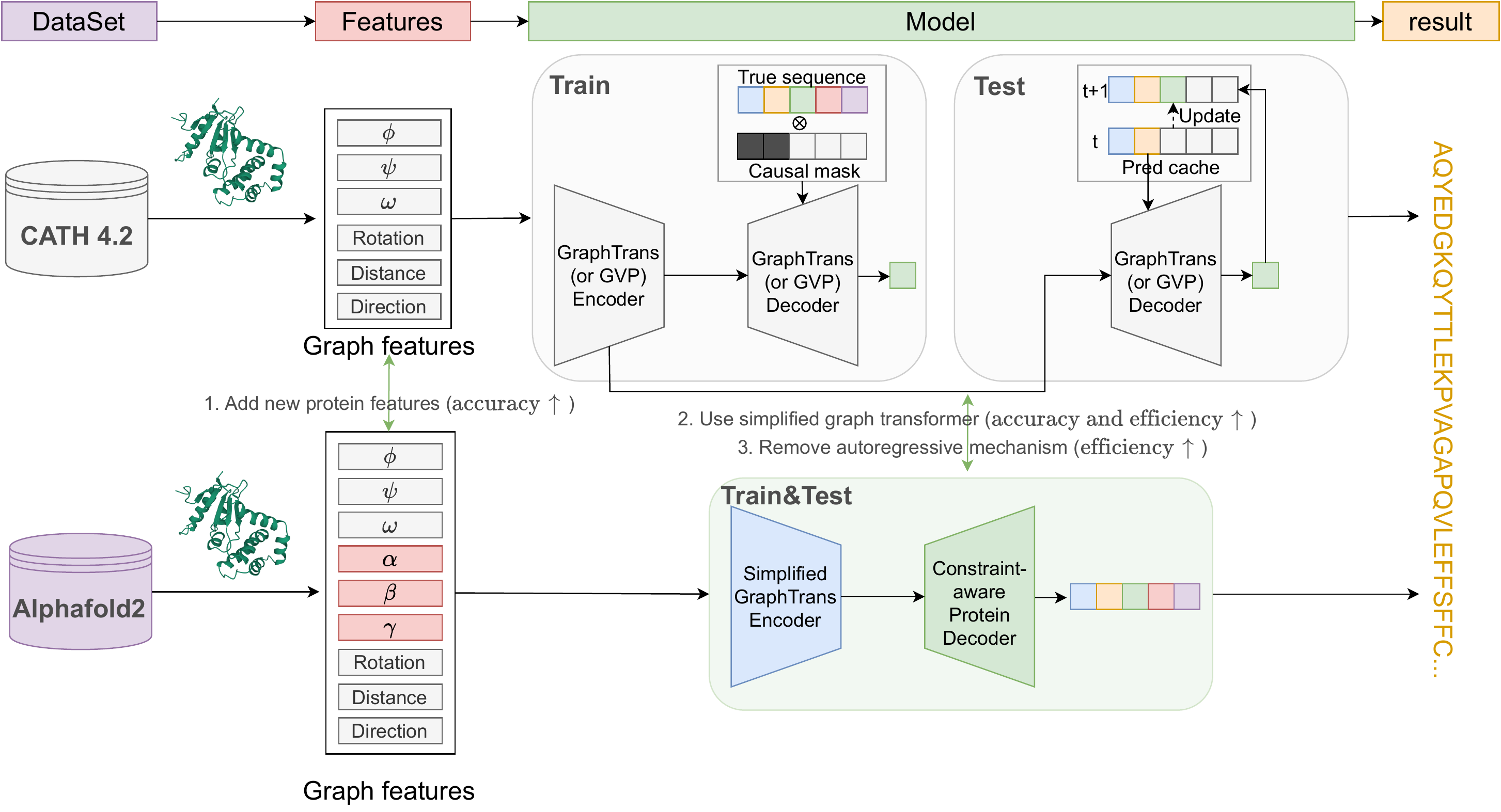}
    \vspace{-2mm}
    \caption{ Overview of ADesign. Compared with GraphTrans, StructGNN \citep{ingraham2019generative} and GVP \citep{jing2020learning}, we add new protein features, simplify the graph transformer, and propose a confidence-aware protein decoder to improve accuracy. }
    \vspace{-2mm}
    \label{fig:overview}
\end{figure*}





\paragraph{Challenges}
There are several challenges that hinder the research progress: (C1) As shown in Table.~\ref{tab:dataset}, different methods may use inconsistent datasets, which will affect the researcher's judgment of the model's ability. (C2) The reported prediction accuracy is still low, i.e., none of the existing models exceeds 50\% accuracy except DenseCPD. (C3) The most accurate CNN model has poor efficiency, e.g.,  DenseCPD takes 7 minutes to predict a 120-length protein on their server. By the way, (C4) the accessibility is poor. Because most protein design models are not open source, subsequent studies become relatively difficult.


\begin{table}[H]
    \centering
    \resizebox{1 \columnwidth}{!}{
    \begin{tabular}{cccccc}
    \toprule
               & Method & $N_{train}$ & $N_{test}$  & Acc & Code\\
    \midrule
    \multirow{3}{*}{\rotatebox{90}{MLP}}    &SPIN       & 1,532    & 500    & 30\% & no  \\
        &SPIN2      & 1,532     & 500   & 34\% & no \\
        &Wang's model    & 10,173   & 50    & 33\% & no \\ \hline
    \multirow{3}{*}{\rotatebox{90}{CNN}}    &SPROF      & 7,134    & 922    & 39.8\% & \href{https://github.com/biomed-AI/SPROF}{PyTorch} \\ 
        &ProDCoNN   & 17,044  & 4,027    & 46.5\% & no\\
        &DenseCPD   & $\leq$10,727  & 500 & 55.53\% & no\\ \hline
    \multirow{2}{*}{\rotatebox{90}{GNN}}    &GraphTrans  & 18,024  & 1,120 & 36.4\% & \href{https://github.com/jingraham/neurips19-graph-protein-design}{PyTorch}\\
        &GVP       & 18,024  & 1,120 & 40.2\% & \href{https://github.com/drorlab/gvp-pytorch}{PyTorch}\\
    \bottomrule
    \end{tabular}}
    \caption{ Statistics of structure-based protein design methods. $N_{train}$ and $N_{test}$ is the number of training set and testing set. Acc indicates the sequence recovery score on TS50.}
    \label{tab:dataset}
\end{table}


For C1, we establish the \textit{new AlphaDesign benchmark} in section.~\ref{sec:experiment}. For C2 and C3, we propose \textit{ADesign method to acheive SOTA accuracy and efficiency}, seeing section.~\ref{sec:method} and section.~\ref{sec:experiment}. As to C4, our model will be open source to accelerate research progress.


\vspace{-2mm}
\section{Methods}
\label{sec:method}





\subsection{Overview}
As shown in Fig.~\ref{fig:overview}, we present the overall framework of ADesign, where the methodological innovations include:



\begin{itemize}
    \vspace{-2mm}
    \item Expressive features: We add new proteins angles $(\alpha, \beta, \gamma)$ to steadily improve the model accuracy.
    \item Better graph encoder: We use the \textbf{s}implified \textbf{g}raph \textbf{t}ransformer (SGT) to extract more expressive representations, where the multi-head attention weights are learned through MLP without multiplying $Q$ and $K$.
    \item Fast sequence decoder: We propose \textbf{c}onstraint-aware \textbf{p}rotein \textbf{d}ecoder (CPD) to replace the autoregressive generator, which can predict residues in parallel and simplify the algorithm.
\end{itemize}

\vspace{-1mm}
With these changes, the proposed ADesign is more accurate, simple, and efficient than previous approaches and, therefore, more attractive as a future baseline model.



\subsection{Graph encoder}
The protein structure can be viewed as a special 3D point cloud in which the order of residues is known. For ordinary 3D points, there are two way to get rotation and translation invariant features: Using special network architecture \citep{fuchs2020se,satorras2021n,jing2020learning,shuaibi2021rotation} that taking 3D points as input or using handicraft invariant features. For proteins, general 3D point cloud approaches cannot consider its particularity, including the regular structure and order of amino acids. Therefore, we prefer learning from the hand-designed, invariant, and protein-specific features. So, which handcrafted invariant features should be constructed? And how to learn expressive representations from these features?






\vspace{-2mm}
\paragraph{Graph}
We represent the protein as a $k$-NN graph derived from residues to consider the 3D dependencies, where $k$ defaults to 30. The protein graph $\mathcal{G}(A,X,E)$ consists of the adjacency matrix  $A \in \{0,1\}^{n,n}$, node features $X \in \mathbb{R}^{n,12}$, and edge features $E \in \mathbb{R}^{m,23}$. Note that $n$ and $m$ are the numbers of nodes and edges, and we create these features by the regular structure, order, and coordinate of residues. 




\begin{figure}[h]
    \centering
    \includegraphics[width=3.25in]{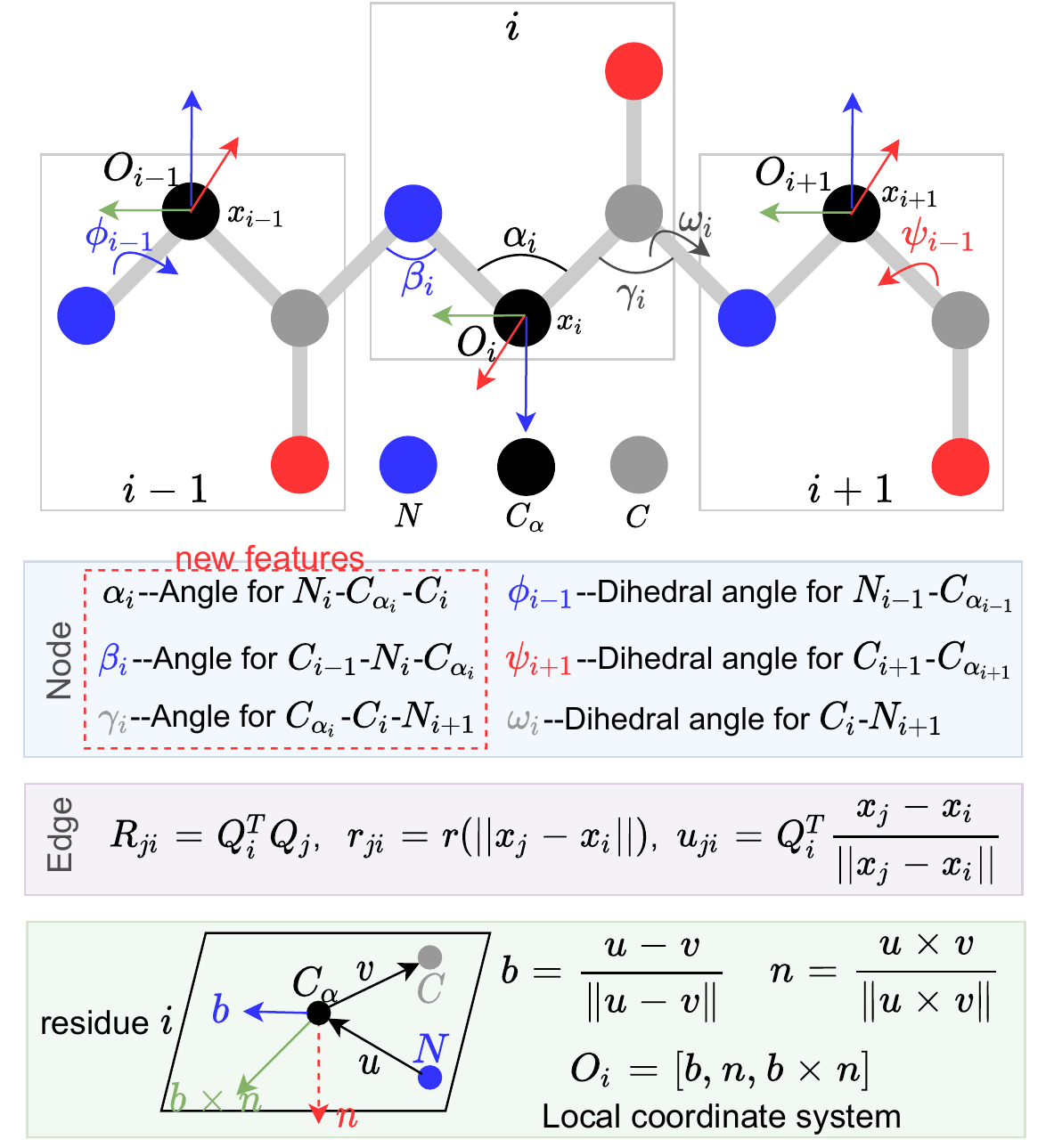}
    \vspace{-3mm}
    \caption{ Angles of the protein backbone.}
    \label{fig:angles}
    \vspace{-3mm}
\end{figure}

\paragraph{Node features} As shown in Fig.~\ref{fig:angles}, we consider two kinds of angles as node features, i.e., the angles $\alpha, \beta, \gamma$ formed by adjacent edges and the dihedral angles $\phi, \psi, \omega$ formed by adjacent surfaces, where $\alpha, \beta, \gamma$ are new features we introduced. For better understanding dihedral angles, it is worth stating that $\phi_{i-1}$ is the angle between plane $C_{i-2}-N_{i-1}-C_{\alpha_{i-1}}$ and $N_{i-1}-C_{\alpha_{i-1}}-C_{i-1}$, whose intersection is $N_{i-1}-C_{\alpha_{i-1}}$. Finally, there are 12 node features derived from $\{ \sin, \cos \} \times \{ \alpha, \beta, \gamma, \phi, \psi, \omega \}$.

\vspace{-2mm}
\paragraph{Edge features} For edge $j \rightarrow i$, we use relative rotation $R_{ji}$, distance $r_{ji}$, and relative orientation $u_{ji}$ as edge features, seeing Fig.~\ref{fig:angles}. At the $i$-th residue's alpha carbon, we establish a local coordinate system $O_i = (b_i,n_i,b_i \times n_i)$. The relative rotation between $O_i$ and $O_j$ is $R_{ji} = Q_i^TQ_j$ can be represented by an quaternion $q(R_{ji})$. We use radial basis $r(\cdot)$ to encode the distance between $C_{\alpha_{i}}$ and $C_{\alpha_{j}}$. The relative direction of $C_{\alpha_j}$ respective to $C_{\alpha_i}$ is calculated by $u_{ji} = Q_i^T \frac{x_j-x_i}{||x_j-x_i||}$. In summary, there are 23 edge features:  4(quaternion) + 16(radial basis) + 3(relative direction).



\begin{figure}[h]
    \centering
    \includegraphics[width=2.5in]{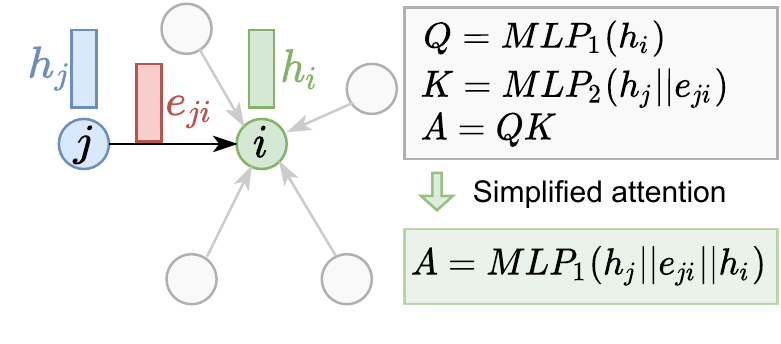}
    \vspace{-3mm}
    \caption{ Simplified graph transformer.}
    \label{fig:attention}
    \vspace{-3mm}
\end{figure}

\vspace{-2mm}
\paragraph{Simplified graph transformer} Denote $\boldsymbol{h}_{i}^{l}$ and $\boldsymbol{e}_{ji}^{l}$ as the output feature vectors of node $i$ and edge $j \rightarrow i$ in layer $l$. We use MLP to project input node and edge features into $d$-dimensional space, thus $\boldsymbol{h}_{i}^{0} \in \mathbb{R}^d$ and $\boldsymbol{e}_{ji}^{0} \in \mathbb{R}^d$. When considering the attention mechanisms centered in node $i$, the attention weight $a_{ji}$ at the $l+1$ layer is calculated by:


\begin{equation}
    \label{eq:attention_weight}
    \begin{cases}
        w_{ji} = MLP_1(\boldsymbol{h}_j^l || \boldsymbol{e}_{ji}^l || \boldsymbol{h}_i^l)\\
        a_{ji} = \frac{\exp{w_{ji}}}{\sum_{k \in \mathcal{N}_i}{\exp{w_{ki}}}}\\
    \end{cases}
\end{equation}

where $\mathcal{N}_i$ is the neighborhood system of node $i$ and $||$ means the concatenation operation. Here, we simplify GraphTrans \citep{ingraham2019generative} by using a single MLP to learn multi-headed attention weights instead of using separate MLPs to learn $Q$ and $K$, seeing Fig.~\ref{fig:attention}. The updated $\boldsymbol{h}_{i}^{l+1}$ is:

\begin{equation}
    \label{eq:update}
    \begin{cases}
        \boldsymbol{v}_j = MLP_2(\boldsymbol{e}_{ji}^l || \boldsymbol{h}_j^l)\\
        \boldsymbol{h}_i^{l+1} = \sum_{j \in \mathcal{N}_i}{a_{ji} \boldsymbol{v}_j}    
    \end{cases}
\end{equation}


By stacking multiple \textbf{s}implified \textbf{g}raph \textbf{t}ransformer (SGT) layers, we can obtain expressive protein representations, which consider both 3D structural constraints by GNN and the 1D information of the manually designed input features.


\begin{figure}[h]
    \centering
    \includegraphics[width=3.3in]{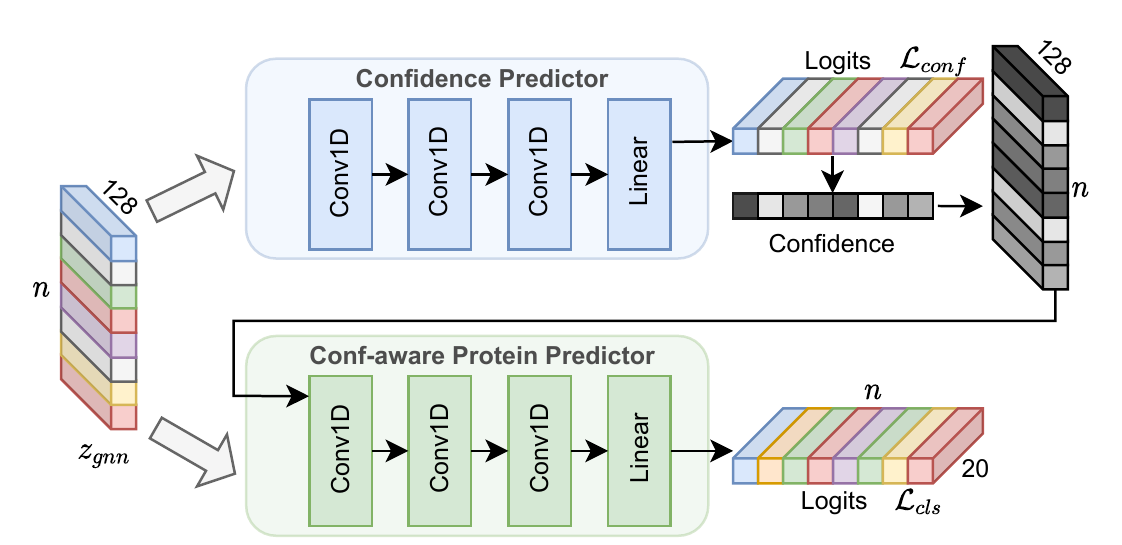}
    \vspace{-3mm}
    \caption{ CPD: The confidence-aware protein decoder. We use two 1D CNN networks to learn confidence scores and make final predictions based on the input graph node features.}
    \label{fig:decoder}
\end{figure}

\vspace{-2mm}
\subsection{Sequence decoder}
To generate more accurate protein sequences, previous researches \citep{ingraham2019generative,jing2020learning} prefer the autoregressive mechanism. However, this technique also significantly slows down the inference process because the residuals must be predicted one by one. \textit{Can we parallelize the predictor while maintaining the accuracy?}




\vspace{-2mm}
\paragraph{Context-aware} Previously, we have considered the 3D constraints through graph network but ignored 1D inductive bias in the neural model. Now, we use 1D CNNs to capture the local sequential dependencies based on the 3D context-aware graph node features. As shown in Fig.~\ref{fig:decoder}, the input features are $\mathcal{Z}_{gnn} = \{z_{1}, z_{2}, \cdots, z_{N}\}$, where $z_{i}$ is the feature vector of node $i$ extracted by previous GNN. The convolution kernel can be viewed as the sliding window. 




\vspace{-2mm}
\paragraph{Confidence-aware} Given the 3D structure $\mathcal{X} = \{x_i: 1 \leq i \leq N\}$ and protein sequence $\mathcal{S} = \{s_i: 1 \leq i \leq N\}$, the vanilla autoregressive prediction indicates $p(\mathcal{S}|\mathcal{X}) = \prod_{i} p(s_i | \mathcal{X}, s_{<i})$, where residues must be predicted one-by-one. We replace autoregressive connections with parallelly estimated confidence score $\boldsymbol{c}$, written as

\vspace{-2mm}
\begin{equation}
    \label{eq:estimator}
    \begin{cases}
    \boldsymbol{c} = f(\text{Conf}(\mathcal{X})) \\
    p(\mathcal{S}|\mathcal{X}) = \prod_{i} p(s_i | \mathcal{X}, x_i, c_{i})
    \end{cases}
\end{equation}
\vspace{-2mm}

where $\boldsymbol{c}$ is the predictive confidence estimated by an auxiliary classifier $\text{Conf}(\cdot)$. The confidence score contains knowledge captured by the previous prediction and plays a similar role to the autoregressive connections. Let $M \in \mathbb{R}^{n,1}$ and $m \in \mathbb{R}^{n,1}$ as Conf's first and secondary largest logits vector of $n$ residues, the confidence score is defined as:

\vspace{-2mm}
\begin{equation}
    \label{eq:confidence}
    \boldsymbol{c} = \lfloor \frac{\boldsymbol{M}}{\boldsymbol{m}} \rfloor .
\end{equation}
\vspace{-2mm}

We encode the confidence score as learnable embeddings $\mathcal{C} \in \mathbb{R}^{n,128}$, concatenate them with graph features, and feed these features into another classifier to get final predictions. Note that all classifiers for estimating confidence and final predictions use the same CE loss:

\vspace{-2mm}
\begin{equation}
    \label{eq:loss}
    \mathcal{L} = - \sum_{i} \sum_{1\leq j \leq 20} \mathbbm{1}_{\{j\}}(y_i) \log(p_{i,j}).
\end{equation}
\vspace{-2mm}

where $p_{i,j}$ is the predicted probability that residue $i$'s type is $j$, $y_i$ is the true label and $\mathbbm{1}_{\{j\}}(\cdot)$ is a indicator function.

    


\section{Experiments}
\label{sec:experiment}
We conduct systemical experiments to establish the new AlphaDesign benchmark and evaluate the proposed ADesgin method. Specifically, we aim to answer:

\begin{itemize}
    \item \textbf{Q1:} What is the difference between the new bechmark and the old one?
    \item \textbf{Q2:} Can ADesign achieve SOTA accuracy on the new benchmark?
    \item \textbf{Q3:} Does ADesign efficient enough in both the training and testing phase?
    \item \textbf{Q4:} What is the secret to achieving SOTA performance? What really matters?
\end{itemize}



\subsection{Benchmark comparation (Q1)}

\paragraph{Metric} Following \citep{li2014direct,o2018spin2,wang2018computational,ingraham2019generative, jing2020learning}, we use sequence recovery to evaluate different protein design methods. Compared with other metrics, such as perplexity, recovery is more intuitive and clear, and its value is equal to the average accuracy of predicted amino acids in a single protein sequence. By default, we report the median recovery score across the entire test set.

\paragraph{Previous benchmark} In Table.~\ref{tab:prev_benchmark}, we show the old benchmark collected from previous studies, including MLP \citep{li2014direct, o2018spin2,wang2018computational}, CNN \citep{chen2019improve,zhang2020prodconn,qi2020densecpd,huang2017densely} and GNN \citep{ingraham2019generative,jing2020learning,strokach2020fast} models. Most approaches report results on the common test set TS50, consisting of 50 small proteins between 60 and 200 in length \citep{li2014direct}. SOTA graph models, e.g., GraphTrans, StructGNN \citep{ingraham2019generative}, and GVP \citep{jing2020learning}, are also compared on a CATH test set with 1120 chains.


\paragraph{Weakness} While the TS50 significantly contributes to establishing the benchmark, we have two concerns: (1) such a small number of proteins does not cover the vast protein space and is more likely to lead to biased tests, and (2) there are no canonical training and validation sets, which means that different methods may use various training sets. Regarding the first concern, TS50 provides limited experimental evidence for protein design, and we need more test data to know how the algorithm performs in more general cases. As to the second concern, if the training data are inconsistent, how can we convince researchers that the performance improvement comes from methods rather than data? Especially when the test set is small, it is much easier to construct the training set to fit the test set manually.



\begin{table}[h]
    \centering
    \resizebox{1.0 \columnwidth}{!}{
    \begin{tabular}{cccccc}
    \toprule
                & $N_{train}$  & CATH  & TS50 \\
    \midrule
    GVP   \citep{jing2020learning}      & 18,024  & 40.2 & 44.9  \\
    StructGNN \citep{ingraham2019generative} & 18,024  & 37.3 & --   \\
    GraphTrans \citep{ingraham2019generative} & 18,024  & 36.4 & --   \\
    DesnseCPD \citep{qi2020densecpd}   & $\leq 10,727$  & --   & 50.7 \\
    ProDCoNN \citep{zhang2020prodconn}   & 17,044   & --           & 40.7 \\
    SBROF  \citep{chen2019improve}       & 7,134    & --           & 39.2 \\
    SPIN2  \citep{o2018spin2}       & 1,532    & --           & 33.6 \\
    Wang's model \citep{wang2018computational}  & 10,179   & --           & 33.0 \\
    ProteinSolver \citep{strokach2020fast} & 18,024    & --          & 30.8 \\
    SPIN    \citep{li2014direct}      & 1,532    & --           & 30.3 \\
    Rosetta \citep{leaver2011rosetta3}      & --    & --            & 30.0 \\
    \bottomrule
    \end{tabular}}
    \caption{Old benchmark. We show recovery scores of various methods on CATH and TS50 test set, where the number of training data $N_{train}$ may vary in different methods.}
    \label{tab:prev_benchmark}
\end{table}


\paragraph{New dataset} We use the AlphaFold Protein Structure Database \footnote{\url{https://alphafold.ebi.ac.uk}} \citep{varadi2021alphafold} to benchmark graph-based protein design methods. As shown in Table.~\ref{tab:alphafoldDB}, there are over 360,000 predicted structures by AlphaFold2 \citep{jumper2021highly} across 21 model-organism proteomes. This dataset has several advantages:
\begin{itemize}
    \setlength{\itemsep}{5pt}
    \setlength{\parsep}{0pt}
    \setlength{\parskip}{0pt}
    \vspace{-2mm}
    \item Species-aware: This dataset provides well-organized proteomic data for different species, which is helpful to develop specialized models for each species.
    
    \item More structures: This dataset provides more than 360,000 structures, while previous Protein Data Bank (PDB) \citep{burley2021rcsb} holds just over 180,000 structures for over 55,000 distinct proteins. 
    

    \item High quality: The median predictive score of AlphaFold2 reaches 92.4\%, comparable to experimental techniques \citep{callaway2020will}. In 2020, the CASP14 benchmark has recognized AlphaFold2 as a solution to the protein–folding problem \citep{pereira2021high}.
    
    \item Missing value: There are no missing values in protein structures provided by AlphaFold DB.
    
    \vspace{-4mm}
\end{itemize}

\begin{table}[t]
    \centering
    \resizebox{1.0 \columnwidth}{!}{
    \begin{tabular}{ccccccc}
    \toprule
    ID & Name   & Structures & $\leq 30$ & $(30,500)$ & $[500,1000]$ & $>1000$\\
    \midrule
    1  & DANRE   &24,664	&27	&15,460	&6,573	&2,604  \\
    2  & CANAL   &5,974	&0	&3,656	&1,874	&444   \\
    3  & MOUSE   &21,615	&55	&13,767	&5,668	&2,125 \\
    4  & ECOLI   &4,363	&63	&3,719	&526	&55     \\
    5  & DROME   &13,458	&28	&8,509	&3,563	&1,358   \\
    6  & METJA   &1,773	&0	&1,605	&144	&24     \\
    7  & PLAF7   &5,187	&1	&2,832	&1,313	&1,041   \\
    8  & MYCTU   &3,988	&4	&3,365	&536	&83     \\
    9  & CAEEL   &19,694	&52	&15,073	&3,592	&977   \\
    10 & DICDI   &12,622	&3	&7,663	&3,367	&1,589   \\
    11 & TRYCC   &19,036	&6	&12,622	&5,007	&1,401  \\
    12 & YEAST   &6,040	&24	&3,789	&1,715	&512    \\
    13 & SCHPO   &5,128	&5	&3,385	&1,386	&352    \\
    14 & RAT   &21,272	&12	&13,884	&5,370	&2,006   \\
    15 & HUMAN   &23,391	&49	&12,399	&5,653	&5,290    \\
    16 & ARATH   &27,434	&42	&19,885	&6,389	&1,118   \\
    17 & MAIZE   &39,299	&83	&29,145	&8,360	&1,711   \\
    18 & LEIIN   &7,924	&0	&4,276	&2,505	&1,143   \\
    19 & STAA8   &2,888	&22	&2,567	&267	&32     \\
    20 & SOYBN   &55,799	&17	&41,048	&12,353	&2,381   \\
    21 & ORYSJ   &43,649	&78	&35,987	&6,738	&846    \\ \hline
       & Sum   &365,198	&571 &254,636	&82,899	&27,092  \\
    \bottomrule
    \end{tabular}}
    \caption{AlphaFold DB: we show the total number of proteins $N_{all}$, and the number of proteins whose length within $(0,30], (30,500], (500,1000]$ and $(1000, +\infty]$. The statistics of all proteomic data are also presented.}
    \label{tab:alphafoldDB}
\end{table}


\paragraph{Novel settings} We extend the experimental setups to length-free and species-aware compared to previous studies \citep{ingraham2019generative,jing2020learning} that use length-limited proteins and do not discriminate species. Length-free means that the protein length may be arbitrary to generalize the model to broader situations; otherwise, the protein must be between 30 and 500 in length. Species-aware indicates that we develop a specific model for each organism's proteome to learn domain-specific knowledge; otherwise, we use the joint proteomic data to train one model for all species. In summary, there are four settings:

\begin{itemize}
    \setlength{\itemsep}{0pt}
    \setlength{\parsep}{0pt}
    \setlength{\parskip}{0pt}
    \item SL: Separate proteomic data + Limited length.
    \item SF: Separate proteomic data + Free length.
    \item JL: Joint proteomic data + Limited length.
    \item JF: Joint proteomic data + Free length.
\end{itemize}

Denote the total amount of structures as $N_{all}$, and the $i$-th species has $N_i$ structures, we have $N_{all} = \sum_{i=1}^{i=21} N_i$. As shown in Table.~\ref{tab:alphafoldDB}, if the length is limited, $N_{all}=254,636$; otherwise, $N_{all}=365,198$. When we use separate proteomic data, we develop 21 models based on datasets with $N_1,N_2,\cdots, N_{21}$ structures; otherwise, we train one model using the joint dataset with $N_{all}$ structures for all species.

\subsection{AlphaDesign benchmark (Q2)}
\paragraph{Data split}   Under the SL and SF setting, for the $i$-th specie, we randomly choose $90\%N_i$ samples for training, $100$ for validation, and $(10\%N_i-100)$ for testing. Correspondingly, in the JL and JF setting, we combine the these sets of all species, resulting in $\sum_{i=1}^{21} 90\%N_i$ proteins for training, $21\times100$ for validation, and $(\sum_{i=1}^{21} 10\%N_i-2100)$ for testing. In addition, it should be noted that for each species, its training sets under SL and SF settings are unrelated, as are the validation sets and test sets. Last but not least, under the same setting, all methods use the same training, validation, and testing sets.

\paragraph{SL and JL settings} In SL and JL settings, the structure length must be between 30 and 500. By default, we use Adam optimizer, OneCycleLR schedular to train all models up to 100 epochs with earlystop patience 20. Besides, we set the bacth\_size of GraphTrans, StructGNN, and ADesign as 16 and the max\_node of GVP as 2000. Note that max\_node represents the maximum number of residuals per batch. Under the SL setting, the learning rate is 0.01. Under the JL setting, we reduce the learning rate and epochs to 0.001 and 50, and increase bacth\_size to 32. Please refer to Table.~\ref{tab:hyperparameter} (in the appendix) for a clearer understanding of the model parameter settings.

\paragraph{SL and JL results} We present the SL and JL benchmarks in Table.~\ref{tab:benchmark_SL}, and conclude that the order of model accuracy from high to low is $\text{ADesign}>\text{GVP}>\text{SGNN}>\text{GTrans}$ under the SL setting. It is worth mentioning that ADesign exceeds previous methods by 8\% accuracy on average. We also observe that ADesign (JL) is less accurate than ADesign (SL), possibly because the latter is more likely to learn species-related knowledge.

\begin{table}[h]
    \resizebox{1.0 \columnwidth}{!}{
    \begin{tabular}{cccccc|c}
    \toprule
          & GTrans & SGNN & GVP  & Our(SL)  & Gain & Our(JL)\\
    \hline
    DANRE & 51.8       & 52.5      & \underline{53.2} & \textbf{63.5}   & +10.3  & 48.2   \\
    CANAL & 45.2       & 46.2      & \underline{48.5} & \textbf{55.1}   & +6.6   & 48.9  \\
    MOUSE & 52.5       & 53.0      & \underline{55.0} & \textbf{64.7}   & +9.7   & 45.7  \\
    ECOLI & 50.0       & 49.4      & \underline{51.9} & \textbf{52.7}   & +0.8   & 50.8 \\
    DROME & 49.6       & 49.2      & \underline{51.2} & \textbf{60.3}   & +9.1   & 47.5  \\
    METJA & 48.9       & 49.6      & \underline{50.9} & \textbf{53.4}   & +2.5   & 48.7  \\
    PLAF7 & 45.0       & 45.9      & \underline{48.6} & \textbf{54.3}   & +5.7   & 37.3  \\
    MYCTU & 53.4       & 53.7      & \underline{56.9} & \textbf{61.8}   & +4.9   & 57.8  \\
    CAEEL & 51.4       & 52.3      & \underline{54.1} & \textbf{63.9}   & +9.8   & 44.1  \\
    DICDI & 49.7       & 50.2      & \underline{52.0} & \textbf{61.4}   & +9.4   & 41.6  \\
    TRYCC & 52.4       & 52.5      & \underline{53.2} & \textbf{65.7}   & +12.5  & 47.9   \\
    YEAST & 44.6       & 44.6      & \underline{48.9} & \textbf{53.9}   & +5.0   & 47.2  \\
    SCHPO & 45.8       & 45.1      & \underline{47.8} & \textbf{53.5}   & +5.7   & 50.4  \\
    RAT   & 52.2       & 52.6      & \underline{53.8} & \textbf{64.2}   & +10.4  & 46.1   \\
    HUMAN & 52.3       & 52.3      & \underline{53.8} & \textbf{63.5}   & +9.7   & 48.2  \\
    ARATH & 52.3       & 53.2      & \underline{54.3} & \textbf{65.1}   & +10.8  & 47.5   \\
    MAIZE & 54.1       & 54.9      & \underline{56.0} & \textbf{67.9}   & +11.9  & 48.3   \\
    LEIIN & 47.6       & 48.8      & \underline{50.7} & \textbf{57.5}   & +6.8   & 54.6  \\
    STAA8 & 46.3       & 46.3      & \underline{48.7} & \textbf{52.4}   & +3.7   & 44.6  \\
    SOYBN & 53.1       & 54.1      & \underline{55.6} & \textbf{67.2}   & +11.6  & 45.1   \\
    ORYSJ & 55.2       & 56.3      & \underline{58.0} & \textbf{69.5}   & +11.5  & 44.9\\ \hline
    Average & 50.2  & 50.6 & \underline{52.5} & \textbf{60.5} & +8.0 & 47.4\\
    \bottomrule
    \end{tabular}}
    \caption{SL and JL benchmarks. We report the median recovery scores across the test set. Due to space constraints, we introduce abbreviations such as GTrans for GraphTrans and SGNN for Struct GNN. ADesign not only compares with baselines in the SL setting, but also provides results under the JL setting. We highlight the best results in bold and underline the sub-optimum ones. }
    \label{tab:benchmark_SL}
\end{table}

\paragraph{SF and JF settings} In SF and JF settings, the protein length can be arbitrary. Under the SF setting, all experimental details remain consistent with the SL setup, except for the bacth\_size of baseline models. Since the maximum structural length can be up to 2700 in AlphaFold DB, we change GVP's max\_node parameter to 3000 to make it applicable to all data. In addition, GraphTrans and StructGNN take more GPU memories because they must pad data according to the longest chain in each batch, and we adjust their batch\_size as 8 to avoid memory overflow. Under the JF setting, we also reduce the learning rate and epochs to 0.001 and 50, and increase bacth\_size to 32. 

\paragraph{SF and JF results} We present the SL and JL benchmarks in Table.~\ref{tab:benchmark_SF} and observe that our ADesign is still the best model, while the sub-optimum results are achieved by SGNN rather than GVP. This probably because GVP is sensitive to the max\_node parameter, while other methods are robust to bacth\_size. Besides, the increase in the amount of training data further improves the model, i.e., the average accuracy of ADesign (SF) is 62.3\% while that of ADesign (SL) is 60.5\%. As a result, ADesign (SF) outperforms baselines by 9.4\%. Finally, we provide results of ADesign (JF) for broader comparations.



 

\begin{table}[h]
    \resizebox{1.0 \columnwidth}{!}{
    \begin{tabular}{cccccc|c}
    \toprule
          & GTrans & SGNN & GVP  & Our(SF) & Gain  & Our(JF) \\
    \midrule
    DANRE & 51.9           & \underline{54.7}      & 44.2 & \textbf{64.2}  & +9.5  & 51.9\\
    CANAL & 48.7           & \underline{51.3}      & 41.3 & \textbf{58.4}  & +7.1  & 53.9\\
    MOUSE & 53.1           & \underline{55.6}      & 44.8 & \textbf{65.6}  & +10.0 & 50.7\\
    ECOLI & \underline{50.6}           & 50.0      & 45.3 & \textbf{56.8}  & +6.2 & 56.0\\
    DROME & 50.2           & \underline{52.3}      & 43.0 & \textbf{62.4}  & +10.1 & 50.6\\
    METJA & \underline{49.8}           & 47.9      & 44.0 & \textbf{53.9}  & +4.1 & 51.1\\
    PLAF7 & 49.6           & \underline{50.6}      & 41.5 & \textbf{58.4}  & +7.8 & 42.5\\
    MYCTU & \underline{53.9}           & 53.2      & 48.5 & \textbf{61.9}  & +8.0 & 59.5\\
    CAEEL & 51.8           & \underline{53.6}      & 43.6 & \textbf{62.6}  & +9.0 & 45.9\\
    DICDI & 52.8           & \underline{54.9}      & 44.9 & \textbf{64.5}  & +9.6 & 47.1\\
    TRYCC & 53.1           & \underline{54.9}      & 45.0 & \textbf{65.9}  & +11.0 & 51.5\\
    YEAST & 47.2           & \underline{47.3}      & 40.5 & \textbf{57.0}  & +9.7 & 51.1\\
    SCHPO & 47.5           & \underline{50.0}      & 41.6 & \textbf{55.7}  & +5.7 & 53.7\\
    RAT   & 53.2           & \underline{55.2}      & 44.2 & \textbf{66.4}  & +11.2 & 49.5\\
    HUMAN & 52.6           & \underline{54.3}      & 44.3 & \textbf{65.1}  & +10.8 & 50.6\\
    ARATH & 53.6           & \underline{55.0}      & 44.6 & \textbf{64.9}  & +9.9 & 50.8\\
    MAIZE & 54.7           & \underline{56.5}      & 47.1 & \textbf{68.7}  & +12.2   & 52.0\\
    LEIIN & 50.9           & \underline{52.2}      & 44.3 & \textbf{63.3}  & +11.1   & 56.9\\
    STAA8 & \underline{46.6}           & 46.3      & 42.5 & \textbf{52.7}  & +6.1   & 50.5\\
    SOYBN & 54.0           & \underline{54.8}      & 45.6 & \textbf{70.3}  & +15.5   & 48.7\\
    ORYSJ & 55.6           & \underline{57.3}      & 46.7 & \textbf{69.1}  & +11.8  & 44.1 \\
    \hline

    Average & 51.5  & \underline{52.8} & 41.2 & \textbf{62.3} & 9.4 & 50.9\\
    \bottomrule
    \end{tabular}}
    \caption{SF and JF benchmarks. We highlight the best recovery scores in bold and underline the sub-optimum ones. }
    \label{tab:benchmark_SF}
\end{table}

\paragraph{Summary} Experiments on both SL and SF settings show that ADesign consistently outperforms previous baseline methods and achieves new SOTA accuracy. This improvement is stable and significant, with an average improvement of over 8\%. If we extend SL and SF to JL and JF, the performance will be degraded, possibly because it is easier to learn species-related knowledge through separated proteomic datasets. However, it is also possible that the insufficient model parameters cause performance degradation. Developing a stronger large model applied to all species could be a future direction.


\vspace{-2mm}
\subsection{Efficiency comparation (Q3)}
A good algorithm should have high accuracy and excellent computational efficiency. Although previous experiments have demonstrated the accuracy advantage of ADesign, we don't know whether it is computationally faster than baselines. Now, we study the training and testing speed of different models on multi-scale datasets.

\vspace{-2mm}
\paragraph{Setting} In both SL and SF settings, we use training data of 5K, 10K, 15K, 20K, and equipped with test samples of 0.5K, 1K, 1.5K, and 2K, respectively, to evaluate the model efficiency. During training, the batch\_size of GraphTrans, StructGNN and ADesign is 8, and the max\_node parameter of GVP is 3000. During testing, each protein is separately fed into the model for prediction, i.e. batch\_size =1.

\begin{figure}[h]
    \centering
    \includegraphics[width=3.3in]{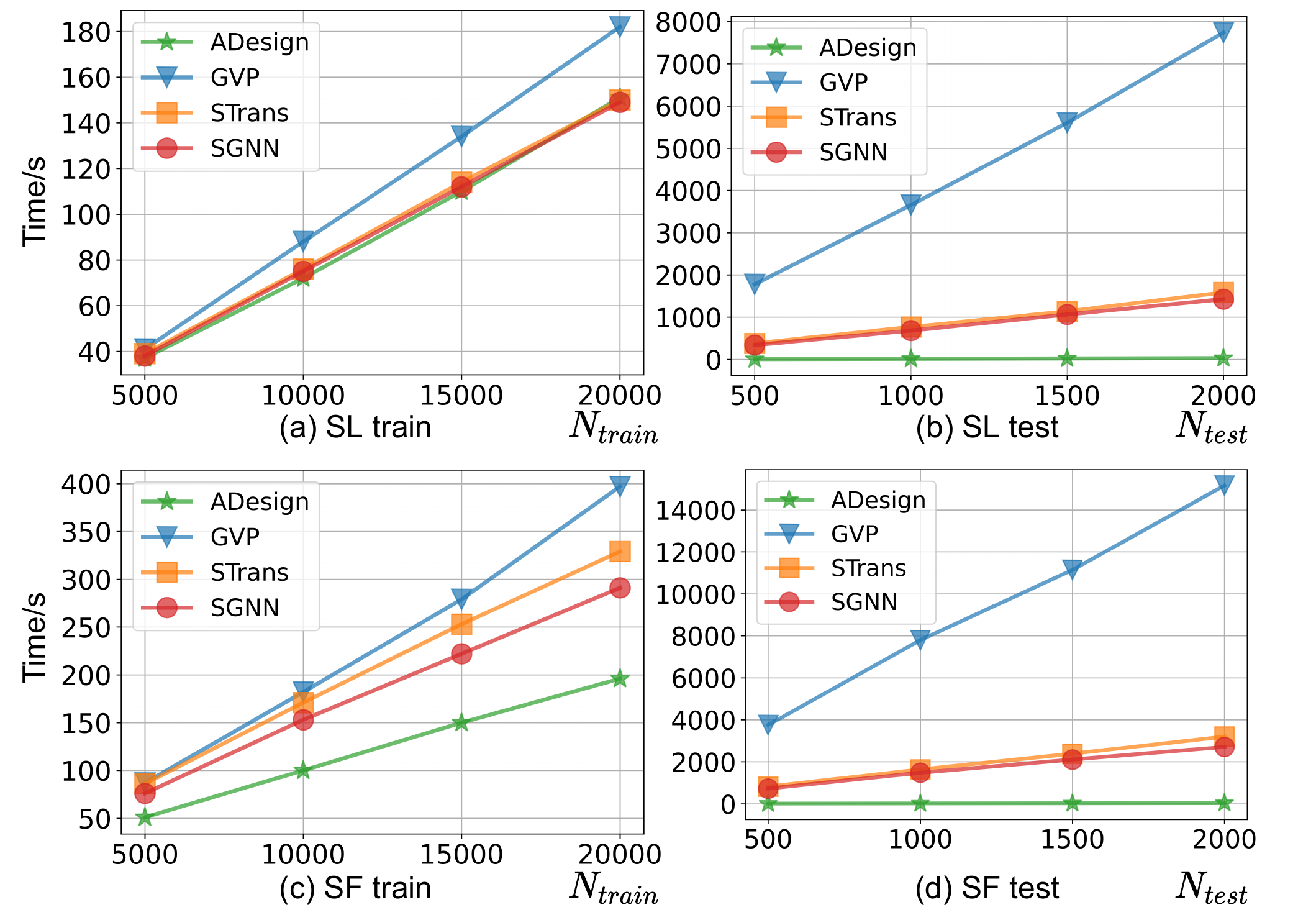}
    \vspace{-4mm}
    \caption{ Speed in SL and SF settings. We show the training time (per epoch) and testing time of various methods, where $N_{train}$ and $N_{test}$ are the data amount of training set and testing set.}
    \label{fig:speed_SL}
\end{figure}

\vspace{-2mm}
\paragraph{Speed} In Fig.~\ref{fig:speed_SL}, we show the running time of various models. In the SL setting, the order of training efficiency from high to low is ADesign$\approx$StructGNN$>$GraphTrans$>$GVP, seeing Fig.~\ref{fig:speed_SL} (a). During testing, ADesign runs at least 40 times faster than other methods, e.g., SGNN takes 1,430s to evaluate 2,000 proteins while ADesign takes only 30 seconds. Under the SF setting, ADesign has the best run-time efficiency at all stages. In contrast, SGNN and GTrans seem to be slower than before, because they need to pad inputs according to the longest structure in each batch, and the longest structure in SL setting is longer than that in SL setting. By the same reasoning, increasing the batch\_size gives ADesign a more significant speedup than SGNN and GTrans at all settings. According to our understanding, the reasons behind the excellent efficiency of ADesign include: (1) We use a simplified graph attention mechanism to speed up feature extraction; (2) We use CPD instead of complex autoregressive decoders to parallelize prediction.

\vspace{-2mm}
\subsection{Ablation study (Q4)}
While ADesign has shown great performance, we are more interested in where the improvements come from. As we have mentioned before, we add new protein features, simplify the graph transformer, and propose a confidence-aware protein decoder to improve accuracy. Then, how much gain can be obtained from each change? Are there redundant changes that are useless for performance improvement?

\vspace{-2mm}
\paragraph{Setting} We conduct ablation experiments on ADesign under the SL setting. Specifically, we may replace the simplified attention module with original GraphTrans (w/o Enc), replace the CPD module with vanilla MLP (w/o Dec), or remove the newly introduced angle features (w/o new feat). Except for model differences, all experimental settings keep the same with previous SL settings.

\vspace{-2mm}
\paragraph{Results and analysis} The ablation results are shown in Table.~\ref{tab:ablation}. Based on the ablation study and section 4.3, we conclude that: (1) The three changes recommended in this paper can bring steady and significant improvements. (2) As to accuracy, the most crucial part is the newly added features, followed by the simplified graph encoder, and finally, the sequence decoder. (3) The confidence-aware protein decoder dramatically improves the evaluation speed, as well as brings considerable accuracy gains.

\begin{table}[h]
    \resizebox{1.0 \columnwidth}{!}{
    \begin{tabular}{cccccc}
    \toprule
           & ADesign  & w/o Enc & w/o Dec  & w/o new feat\\
    \midrule
    DANRE & \textbf{63.5}   & 59.2  & 60.6  & 54.8    \\
    CANAL & \textbf{55.1}   & 49.6  & 52.7  & 42.3   \\
    MOUSE & \textbf{64.7}   & 58.4  & 60.8  & 55.3   \\
    ECOLI & \textbf{52.7}   & 42.5  & 48.4  & 41.9  \\
    DROME & \textbf{60.3}   & 55.1  & 55.8  & 49.7   \\
    METJA & \textbf{53.4}   & 44.0  & 53.2  & 44.1   \\
    PLAF7 & \textbf{54.3}   & 47.5  & 53.4  & 42.6   \\
    MYCTU & \textbf{61.8}   & 55.6  & 57.4  & 50.0   \\
    CAEEL & \textbf{63.9}   & 59.3  & 60.2  & 55.5   \\
    DICDI & \textbf{61.4}   & 55.2  & 58.1  & 50.7   \\
    TRYCC & \textbf{65.7}   & 60.6  & 61.2  & 56.8    \\
    YEAST & \textbf{53.9}   & 48.6  & 52.5  & 42.3   \\
    SCHPO & \textbf{53.5}   & 48.5  & 52.3  & 43.2   \\
    RAT   & \textbf{64.2}   & 58.9  & 61.1  & 54.7    \\
    HUMAN & \textbf{63.5}   & 58.6  & 59.9  & 53.3   \\
    ARATH & \textbf{65.1}   & 61.1  & 62.0  & 55.8    \\
    MAIZE & \textbf{67.9}   & 64.7  & 63.4  & 61.8    \\
    LEIIN & \textbf{57.5}   & 52.3  & 55.8  & 48.0   \\
    STAA8 & \textbf{52.4}   & 43.3  & 52.2  & 43.2   \\
    SOYBN & \textbf{67.2}   & 64.0  & 64.8  & 57.9    \\
    ORYSJ & \textbf{69.5}   & 66.1  & 66.7  & 59.4\\ \hline
    Average & \textbf{60.5} & 54.9 & 57.7 & 50.6\\
    \bottomrule
    \end{tabular}}
    \caption{Ablation study under the SL setting.}
    \label{tab:ablation}
\end{table}

\vspace{-4mm}
\section{Conclusion}
This paper establishes a new benchmark--AlphaDesign, and proposes a new ADesign method for AI-guided protein design. By introducing new protein features, simplifying the graph transformer, and proposing a condifence-aware protein decoder, ADesign achieves state-of-the-art accuracy and efficiency. We hope this work will help standardize comparisons and provide inspiration for subsequent researches.

\bibliography{AlphaDesign}
\bibliographystyle{icml2021}

\onecolumn
\appendix
\section{Appendix}

\subsection{Hyper-parameters}


\begin{table*}[h]
    \resizebox{1.0 \columnwidth}{!}{
    \begin{tabular}{ccccccccccc}
        \toprule
    Setting   & \begin{tabular}[c]{@{}c@{}}Seperate\\ data\end{tabular} & \begin{tabular}[c]{@{}c@{}}Limited \\ length\end{tabular} & $N_{train}$ & $N_{val}$ & $N_{test}$ & \begin{tabular}[c]{@{}c@{}}batch\_size\\ (GraphTrans)\end{tabular} & \begin{tabular}[c]{@{}c@{}}max\_node\\ (GVP)\end{tabular} & \begin{tabular}[c]{@{}c@{}}batch\_size\\ (ADesign)\end{tabular} & lr    & epoch \\
     \midrule
    SL & yes  & yes   & $0.9 N_i$     & 100  & $0.1 N_i - 100$     &16  & 2000  & 16    & 0.01  & 100   \\
    SF & yes  & no    & $0.9 N_i$     & 100  & $0.1 N_i - 100$     &8  & 3000  & 16    & 0.01  & 100   \\
    JL & no   & yes   & $0.9 N_{all}$ & 100  & $0.1 N_{all} - 100$ & -- & --    & 32    & 0.001 & 50    \\
    JF & no   & no    & $0.9 N_{all}$ & 100  & $0.1 N_{all} - 100$ & -- & --    & 32    & 0.001 & 50   \\
    \bottomrule
    \end{tabular}}
    \caption{Hyper-parameters. $N_i$ is the number of strcutures in the $i$-th species' proteome, and $N_{all}$ is the number of structures of all species. And max\_node plays the similar role as batch\_size, which represents the maximum number of residuals per batch.}
    \label{tab:hyperparameter}
\end{table*}

\subsection{Invariance}
\paragraph{Experimental details} For GraphTrans, StructGNN and GVP, we use their open-source PyTorch code without any changes. The platform for our experiment is ubuntu 18.04, with 8 Intel(R) Xeon(R) Gold 6240R Processors and 256GB memory. We use a single NVIDIA V100 to train and evaluate all models, where the CUDA version is 11.3 and Torch version is 1.10.1. To evaluate invariance, we retest pre-trained models (SL setting) on the test set that has been randomly rotated and translated. Besides, we provide the inference time of each model.

\paragraph{Results and analysis} As shown in Table.~\ref{tab:invariance}, we conclude that GTrans, SGNN and ADesign are rotationally and translationally invariant considering the computational uncertainty. This property is determined at the feature level: all input features are invariant to rotations and translations. In contrast, GVP is equivalent, but not invariant, while the latter is what we really need for classification. In addition, ADesign is much faster than other methods during inference. However, the reported time includes model building and data loading and is not purely runtime, so it does not seem to be 40 times faster than the others.

\begin{table}[h]
    \resizebox{1.0 \columnwidth}{!}{
    \begin{tabular}{ccccc|cccc|cccc}
        \toprule
            & \multicolumn{4}{c|}{Raw data}   & \multicolumn{4}{c|}{Rotation+Translation} & \multicolumn{4}{c}{Inference time}   \\
            & GTrans & SGNN & GVP  & ADesign & GTrans    & SGNN    & GVP    & ADesign   & GTrans & SGNN   & GVP      & ADesign \\
        \hline
    DANRE   & 51.8   & 52.5 & \underline{53.2} & \textbf{63.5}    & 51.9      & \underline{52.5}    & 42.6   & \textbf{63.7}      & 21m30s & \underline{19m37s} & 1h30m58s & \textbf{37s  }   \\
    CANAL   & 45.2   & 46.2 & \underline{48.5} & \textbf{55.1}    & 45.6      & \underline{46.0}    & 40.5   & \textbf{55.1}      & 4m9s   & \underline{3m48s } & 17m35s   & \textbf{13s  }   \\
    MOUSE   & 52.5   & 53.0 & \underline{55.0} & \textbf{64.7}    & 52.4      & \underline{53.1}    & 44.5   & \textbf{64.7}      & 19m6s  & \underline{18m1s } & 1h18m42s & \textbf{37s  }   \\
    ECOLI   & 50.0   & 49.4 & \underline{51.9} & \textbf{52.7}    & 49.5      & \underline{50.0}    & 44.0   & \textbf{52.5}      & 3m40s  & \underline{3m18s } & 14m54s   & \textbf{18s  }   \\
    DROME   & 49.6   & 49.2 & \underline{51.2} & \textbf{60.3}    & 49.7      & \underline{49.6}    & 43.0   & \textbf{60.2}      & 10m14s & \underline{9m33s } & 45m19s   & \textbf{24s  }   \\
    METJA   & 48.9   & 49.6 & \underline{50.9} & \textbf{53.4}    & 48.4      & \underline{49.3}    & 43.1   & \textbf{54.0}      & 54s    & \underline{49s   } & 3m21s    & \textbf{10s  }   \\
    PLAF7   & 45.0   & 45.9 & \underline{48.6} & \textbf{54.3}    & 45.1      & \underline{45.8}    & 40.9   & \textbf{54.0}      & 2m48s  & \underline{2m27s } & 11m12s   & \textbf{12s  }   \\
    MYCTU   & 53.4   & 53.7 & \underline{56.9} & \textbf{61.8}    & 53.3      & \underline{53.7}    & 47.1   & \textbf{61.6}      & 3m17s  & \underline{2m56s } & 13m18s   & \textbf{13s  }   \\
    CAEEL   & 51.4   & 52.3 & \underline{54.1} & \textbf{63.9}    & 51.7      & \underline{52.1}    & 44.6   & \textbf{63.9}      & 19m41s & \underline{17m48s} & 1h22m29s & \textbf{39s  }   \\
    DICDI   & 49.7   & 50.2 & \underline{52.0} & \textbf{61.4}    & 49.4      & \underline{50.1}    & 43.6   & \textbf{61.1}      & 8m47s  & \underline{8m0s  } & 36m29s   & \textbf{26s  }   \\
    TRYCC   & 52.4   & 52.5 & \underline{53.2} & \textbf{65.7}    & 52.1      & \underline{52.2}    & 43.8   & \textbf{65.7}      & 17m31s & \underline{14m40s} & 1h11m27s & \textbf{33s  }   \\
    YEAST   & 44.6   & 44.6 & \underline{48.9} & \textbf{53.9}    & 44.6      & \underline{44.8}    & 39.5   & \textbf{53.8}      & 3m49s  & \underline{3m30s } & 15m39s   & \textbf{14s  }   \\
    SCHPO   & 45.8   & 45.1 & \underline{47.8} & \textbf{53.5}    & \underline{45.9}      & 45.1    & 39.6   & \textbf{53.6}      & 3m29s  & \underline{3m11s } & 14m49s   & \textbf{17s  }   \\
    RAT     & 52.2   & 52.6 & \underline{53.8} & \textbf{64.2}    & 52.0      & \underline{52.6}    & 44.2   & \textbf{64.4}      & 19m14s & \underline{17m58s} & 1h17m13s & \textbf{39s  }   \\
    HUMAN   & 52.3   & 52.3 & \underline{53.8} & \textbf{63.5}    & 52.2      & \underline{52.3}    & 44.9   & \textbf{63.6}      & 17m24s & \underline{15m15s} & 1h10m14s & \textbf{33s  }   \\
    ARATH   & 52.3   & 53.2 & \underline{54.3} & \textbf{65.1}    & 52.3      & \underline{53.2}    & 45.0   & \textbf{65.1}      & 28m42s & \underline{24m27s} & 1h55m27s & \textbf{57s  }   \\
    MAIZE   & 54.1   & 54.9 & \underline{56.0} & \textbf{67.9}    & 54.1      & \underline{54.8}    & 43.7   & \textbf{67.8}      & 39m25s & \underline{33m25s} & 2h43m43s & \textbf{1m11s}   \\
    LEIIN   & 47.6   & 48.8 & \underline{50.7} & \textbf{57.5}    & 47.8      & \underline{48.5}    & 42.8   & \textbf{57.5}      & 5m19s  & \underline{4m53s } & 21m54s   & \textbf{15s  }   \\
    STAA8   & 46.3   & 46.3 & \underline{48.7} & \textbf{52.4}    & \underline{46.0}      & \underline{46.0}    & 39.7   & \textbf{52.7}      & 2m1s   & \underline{1m45s } & 7m47s    & \textbf{12s  }   \\
    SOYBN   & 53.1   & 54.1 & \underline{55.6} & \textbf{67.2}    & 53.1      & \underline{54.0}    & 44.7   & \textbf{67.2}      & 51m35s & \underline{50m12s} & 3h41m53s & \textbf{1m41s}   \\
    ORYSJ   & 55.2   & 56.3 & \underline{58.0} & \textbf{69.5}    & 55.1      & \underline{56.3}    & 42.2   & \textbf{69.5}      & 40m58s & \underline{35m15s} & 2h52m11s & \textbf{1m20s}   \\ \hline
    Average & 50.2   & 50.6 & \underline{52.5} & \textbf{60.5}    & 50.1      & \underline{50.6}    & 43.0   & \textbf{60.6}      & --     & --                 & --       & --  \\
    \bottomrule
    \end{tabular}}
    \caption{Invariance. We retest pre-trained models (SL setting) on the test set that has been randomly rotated and translated. We also provide the inference time that includes model building and data loading.}
    \label{tab:invariance}
    \end{table}

\end{document}